# Optical properties of $CdP_2$ and $CdP_4$ nanoparticles incorporated into Na-X zeolite matrix

O.A.Yeshchenko [a,*], I.M.Dmitruk [a], S.V.Koryakov [a] and Yu.A.Barnakov [b]

[a] *Physics Department, National Taras Shevchenko Kyiv University,*
*6 Akademik Glushkov prosp., 03127 Kyiv, Ukraine*
[b] *Advanced Materials Research Institute, University of New Orleans, New Orleans, LA 70148*

For the first time nanoparticles of $CdP_2$ and $CdP_4$ semiconductors were prepared and studied. The nanoparticles were prepared by incorporation into zeolite Na-X matrix. Absorption and photoluminescence (PL) spectra of $CdP_2$ and $CdP_4$ nanoparticles were measured at the temperature of 77 K. Two bands were observed in both the absorption and PL spectra of $CdP_2$ nanoparticles demonstrating the blue shift from the line of free exciton in bulk crystal. Single band was observed in both the absorption and PL spectra of $CdP_4$ nanoparticles demonstrating the blue shift from the spectra of bulk crystal. We attribute the bands to some stable nanoparticles of $CdP_2$ and $CdP_4$ with size less than the size of zeolite Na-X supercage. We observed Stokes shift of the PL bands with respect to the absorption bands for both $CdP_2$ and $CdP_4$ nanoparticles.

## 1. Introduction

Many different methods have been used for preparation of the semiconductor nanoparticles, e.g. preparation of nanoparticles in solutions [1], glasses [2] or polymers [3]. However, it is not easy to control the size distribution of small particles with countable number of atoms in these methods. Matrix method based on the incorporation of materials into the 3D regular system of voids and channels of zeolites crystals could be one of the possible solution [4,5]. Zeolites are crystalline alumosilicates with cavities which size can vary in the range from one to several tents nanometers. It depends on the type of alumosilicate framework, ratio Si/Al, origin of ion-exchanged cations, which stabilise negative charge of framework, etc. Zeolite NaX, which has been used in the present work has Si/Al ratio equal 1, Fd3m symmetry and two types of cages: one is sodalite cage – truncated octahedron with diameter 8 Å and supercage, which is formed by the connection of sodalites in diamond-like structure with the diameter of about 13 Å [6]. All cages are interconnected by shared small windows and arranged regularly. Thus, the cages can be used for preparion of small semiconductor nanoparticles.

---
[*] Corresponding author: O.A.Yeshchenko.
Tel.: +380-44-2664587;   Fax: +380-44-2664036;   E-mail: yes@univ.kiev.ua

To the best of our knowledge the first report of the preparation and study of II-V semiconductor nanoparticles is our recent work [7] where the optical properties of the nanoparticles of zinc diphosphide ($ZnP_2$) incorporated into zeolite Na-X matrix have been studied. The present paper is the first study of other II-V nanoparticles: cadmium diphosphide ($CdP_2$) and cadmium tetraphosphide ($CdP_4$). Wet chemistry methods are not applicable for a production of ultrasmall II-V nanoparticles due to their high reactivity in water. It is hard to expect their stability in glass melt as well. Thus, incorporation into zeolite cages seems to us to be one of the most suitable methods of preparation of II-V semiconductor nanoparticles ($CdP_2$ and $CdP_4$ in particular).

Quantum confinement of charge carriers in small particles leads to new effects in optical properties of the particles. Those are the blue shift of exciton spectral lines originating from the increase of the kinetic energy of charge carriers and the increase of the oscillator strength per unit volume [8,9]. These effects are quite remarkable when the size of the nanoparticle is comparable with or smaller than Bohr radius of exciton in bulk crystal, i.e. when the regime of the strong confinement is realised. Since zeolite cages are rather small and can contain only small nanoparticles, the incorporation of semiconductor nanoparticles into the zeolite matrix is the easy way to realise the strong confinement regime for electrons and holes.

Bulk $CdP_2$ crystal is the indirect-gap semiconductor (energy gap: 2.155 eV: see, e.g. Ref. [10]). The symmetry of its lattice is characterised by the space symmetry group $D_4^4$ for right-rotating and $D_4^8$ for left-rotating modification (tetragonal syngony). As the bulk crystal is red at the temperature of 77 K, the blue shifted exciton lines of $CdP_2$ nanoparticles are expected to be most probably in visible or near UV spectral region.

Bulk $CdP_4$ crystal is the direct-gap semiconductor (energy gap: 0.908 eV: see Ref. [11]). The symmetry of its lattice is characterised by the space symmetry group $C_{2h}^5$ (monoclinic syngony). As the bulk crystal has so narrow energy gap, the blue shifted exciton lines of $CdP_4$ nanoparticles are expected to be most probably in the visible or near IR spectral region.

## 2. Technological and experimental procedures

For the preparation of $CdP_2$ and $CdP_4$ nanoparticles we used crystals of $CdP_2$ and $CdP_4$ and synthetic zeolite of Na-X type. The framework of zeolite Na-X consists of sodalite cages and supercages with the inside diameters of 8 and 13 Å, respectively. Both $CdP_2$ molecule and $CdP_4$ one seem to us to be too large to be incorporated into small sodalite cage, because of the existence of many Na cations. Therefore, it is naturally to assume that only the supercages can be the hosts for the nanoparticles. Zeolite, $CdP_2$ and $CdP_4$ crystals were dehydrated in quartz ampoule in vacuum about $2\times10^{-5}$ mm Hg for 1 h at 400°C. Then ampoule was sealed. We used 100 mm length ampoule for space separation of semiconductor source and zeolite in it. The preparation of samples of both compounds was carried in two stages. At the first stage $CdP_2$ was incorporated into the zeolite matrix through the vapour phase at 774°C in source region and 760°C in zeolite region for 75 h. Respectively, $CdP_4$ was incorporated into the zeolite matrix through the vapour phase at 676°C in source region and 671°C in zeolite region for 75 h. At the second stage, the inverted temperature gradient was applied: for $CdP_2$ – 745°C in source region and 765°C in zeolite region, for $CdP_4$ – 657°C in source region and 662°C in zeolite region. The duration of the second stage was 50 h. The cooling of ampoule we carried out gradually with above mentioned inverted temperature gradient. The stability of structure of lattice of the zeolite monocrystals was controlled by XRD method. The control showed that at above mentioned temperatures the zeolite lattice structure was stable, i.e. semiconductor nanoparticles were incorporated into the monocrystal zeolite matrix.

Samples in quartz ampoule were dipped into liquid nitrogen during the experiment. A tungsten-halogen incandescent lamp was used as a light source for the diffuse reflection measurements. An $Ar^{++}$ laser with wavelength 3511 Å was used for the excitation of the luminescence of $CdP_2$ nanoparticles, and $Ar^+$ laser with wavelength 4880 Å – for PL of $CdP_4$ ones.

In present work, the absorption spectra of the nanoparticles were obtained from the diffuse reflection spectra by conversion with Kubelka-Munk function $K(\hbar\omega)=[1-R(\hbar\omega)]^2/2R(\hbar\omega)$, where $R(\hbar\omega)$ is the diffuse reflectance normalised by unity at the region of no absorption.

### 3. Optical properties of $CdP_2$ nanoparticles

Diffuse reflection (DR) and photoluminescence (PL) spectra of the $CdP_2$ nanoparticles incorporated into the 13 Å supercages of zeolite Na-X were measured at the temperature of 77 K. Then, the DR spectrum was converted to absorption one by the Kubelka-Munk method described above. The obtained absorption spectrum is presented in Fig. 1(a). The spectrum demonstrates clear two-band structure. The spectral positions of the respective bands signed as $C_1$ and $C_2$ are presented in Table 1. Both the bands demonstrate blue shift (Table 1) from the spectrum of the bulk crystal. Let us note that the blue shift of the absorption bands for $CdP_2$ nanoparticles in Na-X zeolite is larger than one for $ZnP_2$ nanoparticles in the same zeolite [7]: 0.628-1.127 eV for $CdP_2$ and 0.282-0.808 eV for $ZnP_2$. The observed blue shift allows us to attribute these bands to the absorption into exciton states of $CdP_2$ nanoparticles incorporated into supercages of zeolite. The photoluminescence spectrum of $CdP_2$ nanoparticles (Fig. 1(b)) shows the same structure as the absorption one, i.e. PL spectrum consists of the corresponding two $C_1'$ and $C_2'$ bands. Their spectral positions are presented in Table 1. It is seen that PL bands of nanoparticles are blue shifted from the luminescence bands of bulk crystal as well (Table 1). The observed blue shift of the absorption and luminescence bands is the result of the quantum confinement of electrons and holes in $CdP_2$ nanoparticles. As the exciton Bohr radius in bulk crystal is larger than radius of zeolite supercage, the strong confinement regime takes place in the nanoparticles.

It is often observed that nanoparticles with certain number of atoms are characterised by higher binding energy (stable nanoparticles) and are more abundant in the sample. This effect is well known for the nanoparticles of different types, e.g. for C [12], Ar [13], Na [13], and for nanoparticles of II-VI semiconductors [14,15]. Our ab initio calculations [7] have shown that such stable nanoparticles exist for $ZnP_2$. Those are $(ZnP_2)_6$ and $(ZnP_2)_8$ with binding energies 1.90 eV and 2.12 eV per atom respectively. So, one can assume the existence of such stable $(CdP_2)_n$ nanoparticles for $CdP_2$ as well. Now, we perform the ab initio calculations aimed to find such stable $CdP_2$ particles. Thus, $C_1$ and $C_2$ bands can be attributed to two stable $CdP_2$ nanoparticles incorporated into the supercages of zeolite matrix. Probably, they are stoichiometric since we do not have any evidence of $CdP_2$ decomposition at the temperature used for its sublimation.

One can see from the Table 1 that the luminescence bands have the Stokes shift from the absorption ones. The values of this shift are large enough: 0.758 eV for $C_1$ band and 0.521 eV for $C_2$ one. These values are considerably larger than ones for $ZnP_2$ nanoparticles in the same Na-X zeolite (0.078-0.135 eV: see Ref. [7]). Stokes shift is well known both in the molecular spectroscopy and in the spectroscopy of nanoparticles. It is known that this kind of Stokes shift (so-called Frank-Condon shift) is due to vibrational relaxation of the excited molecule or nanoparticle to the ground state. It is seen that for smaller nanoparticle ($C_1$-band) the Stokes shift is larger. Such dependence can be explained by the theory developed, e.g. in Ref. [16] where the first-principle calculations of excited-state relaxations in nanoparticles and the dependence of respective Stokes shift on particle size were performed. As it is shown in Ref. [16], for small nanoparticles the Stokes shift is the Frank-Condon one, which is the result of the vibrational relaxation of the nanoparticle in the excited electronic state.

## 4. Optical properties of $CdP_4$ nanoparticles

Absorption and photoluminescence spectra of the $CdP_4$ nanoparticles incorporated into the 13 Å supercages of zeolite Na-X measured at the temperature of 77 K are shown in Fig. 3. The absorption spectrum demonstrates single-band structure. The spectral position of the respective $D_1$ band is presented in Table 1. The band demonstrates the blue shift (Table 1) from the spectrum of bulk crystal. The blue shift of the absorption band for $CdP_4$ nanoparticles is 1.076 eV that is close to respective value for $CdP_2$ nanoparticles (0.628-1.127 eV) and is larger than one for $ZnP_2$ nanoparticles (0.282-0.808 eV: see Ref. [7]) in zeolite of the same type. The observed blue shift allows us to attribute this band to the absorption into exciton states of $CdP_4$ nanoparticles incorporated into supercages of zeolite. The photoluminescence spectrum of $CdP_4$ nanoparticles (Fig. 3) shows the same structure as the absorption one, i.e. PL spectrum consists of single $D_1'$ band. Its spectral position is presented in Table 1. The PL band of nanoparticles is characterised by blue shift from the spectrum of bulk crystal as well (Table 1). The observed blue shift of the absorption and luminescence bands is the result of the quantum confinement of electrons and holes in $CdP_4$ nanoparticles. As the exciton Bohr radius in bulk crystal is larger than radius of zeolite supercage, the strong confinement regime takes place in $CdP_4$ nanoparticles.

Similarly to the $CdP_2$ nanoparticles, one can assume that the $D_1$ band corresponds to the stable nanoparticle. Since, there is single band in the spectra, we can assume that only single type of the stable $CdP_4$ nanoparticle exists in the radius region up to 13 Å. Probably, they are stoichiometric since we do not have any evidence of $CdP_4$ decomposition at the temperature used for its sublimation. Now, we perform the theoretical search of the stable $CdP_4$ nanoparticles.

One can see from the Table 1 that the luminescence band of $CdP_4$ nanoparticles has the Stokes shift from the absorption one. The value of this shift (0.562 eV) is quite large and it is close to Stokes shift for $CdP_2$ nanoparticles (0.521-0.758 eV). But, it is considerably larger than one for $ZnP_2$ nanoparticles (0.078-0.135 eV: see Ref. [7]) in the same Na-X zeolite. Evidently, similar to $CdP_2$ nanoparticles the Stokes shift for $CdP_4$ ones is the Frank-Condon shift, which is the result of the vibrational relaxation of the nanoparticle in the excited electronic state.

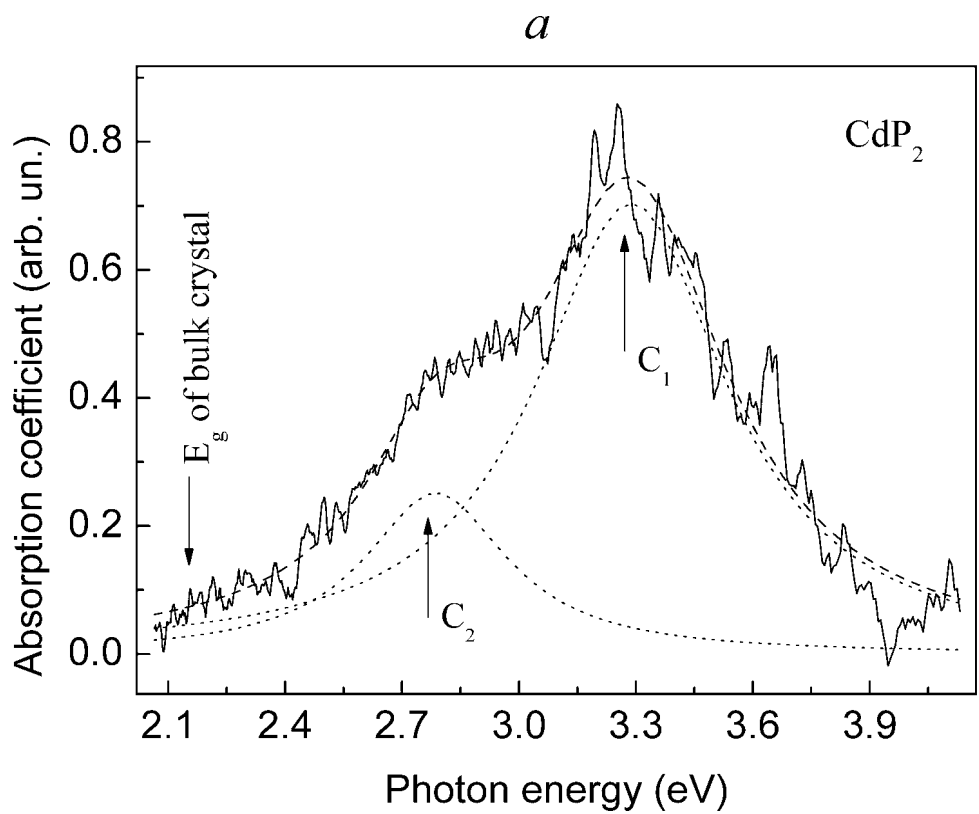

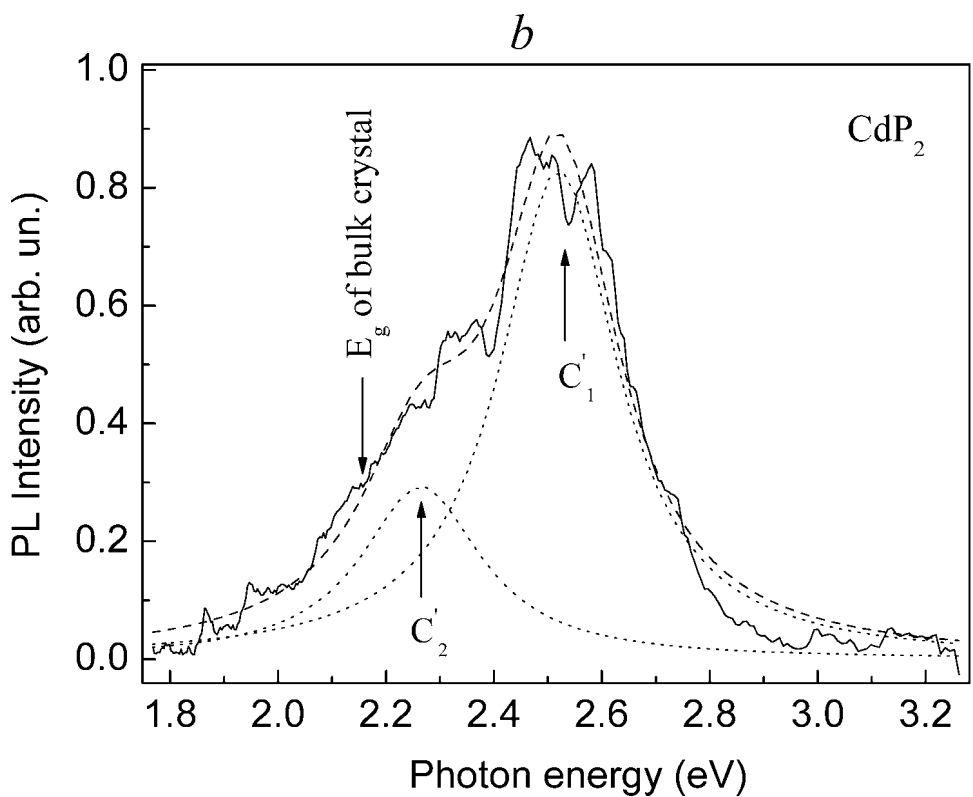

Fig. 1. (a) – The obtained by Kubelka-Munk method absorption spectrum and (b) – the photoluminescence spectrum of $CdP_2$ nanoparticles in zeolite Na-X at the temperature of 77 K.

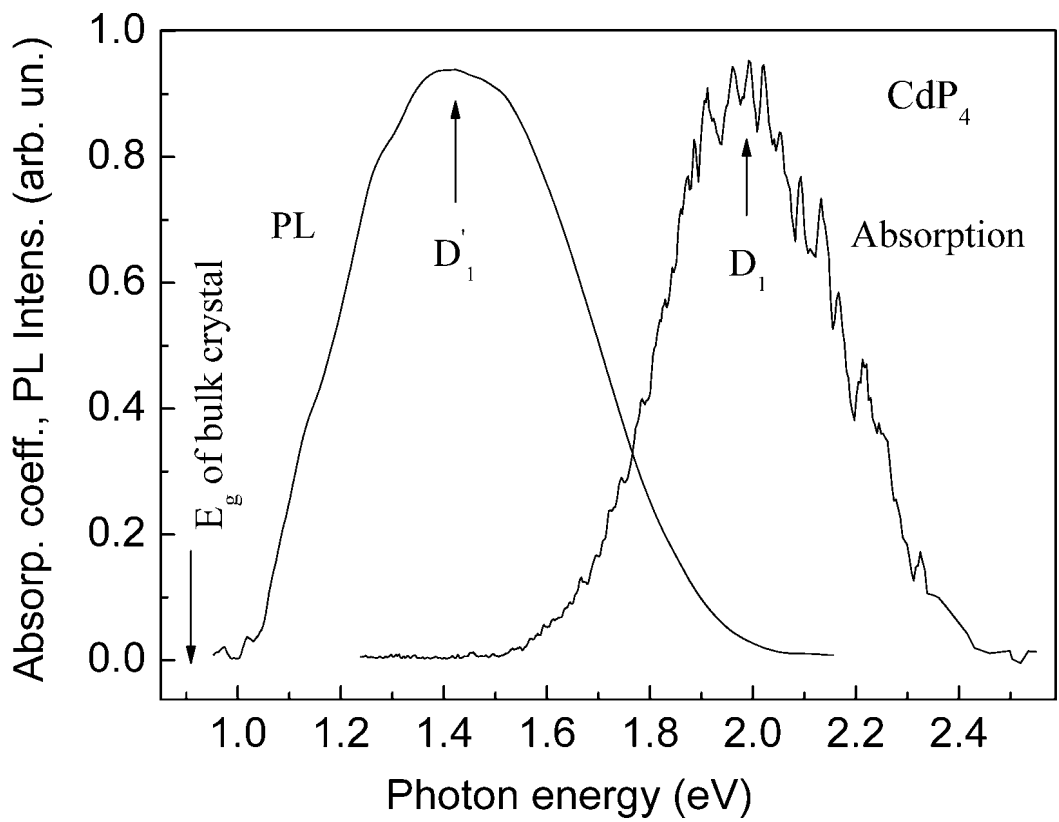

Fig. 2. The obtained by Kubelka-Munk method absorption spectrum and the photoluminescence spectrum of $CdP_4$ nanoparticles in zeolite Na-X at the temperature of 77 K.

Table 1. Spectral characteristics of $CdP_2$ and $CdP_4$ nanoparticles in zeolite Na-X.

| Substance | Spectral position (eV) | | Blue shift of absorption bands (eV) | Stokes shift (eV) |
|---|---|---|---|---|
| | Absorption | PL | | |
| $CdP_2$ | 3.282 ($C_1$) | 2.524 ($C_1'$) | 1.127 | 0.758 |
| | 2.783 ($C_2$) | 2.262 ($C_2'$) | 0.628 | 0.521 |
| $CdP_4$ | 1.984 ($D_1$) | 1.422 ($D_1'$) | 1.076 | 0.562 |